\documentclass[useAMS,usenatbib,twocolumn]{mn2e}
\usepackage{graphicx,times}
\usepackage[usenames,dvipsnames]{color}
\usepackage{amsmath,amssymb,bm}
\usepackage{etoolbox}
\def \ff {{\bm f}}
\newcommand{\eee}{\hat{\bm e}}
\def \fh {\tilde{{\bm f}}}

\def \xx {{\bm x}}
\def \kk {{\bm k}}
\def \ii {{\rm i}}
\def \wf {w_{\rm f}}
\def \erf {{\rm erf}}
\def \smax {\sigma_{\rm max}}
\newcommand{\RRRR}{\mbox{\boldmath ${\sf R}$} {}}

\def \del2z {\partial^{2}_{z}}

\def \Bhx {B^{\rm h}_{x}}

\def \Beq {B_{\rm eq}}
\def \Beqz {B^0_{\rm eq}}

\def \AAA {{\bm A}}
\def \ww {{\bm w}}
\def \WW {{\bm W}}
\def \uu {{\bm u}}
\def \UU {{\bm U}}

\def \Wbar {{\overline {\bm W}}}
\def \Ubar {{\overline {\bm U}}}

\def \JJ {{\bm J}}
\newcommand{\SSSS}{\mbox{\boldmath ${\sf S}$} {}}

\def \grav {{\bm g}}
\def \grad {{\bm \nabla}}
\def \curl {{\bm \nabla} \times}

\def \delt {\partial_t}
\def \Dt {D_t}

\newcommand{\brah}[1]{\left\langle #1\right\rangle_{xy}}
\newcommand{\braf}[1]{\left\langle #1\right\rangle_{\kappa=5}}
{}

\def \Rm  {\mbox{Re}_{\rm M}}

\def \Rey  {\mbox{Re}}

\def \Tto  {\tau_{\rm to}}
\def \Teta  {\tau_{\rm td}}

\def \kf  {k_{\rm f}}

\def \urms  {u_{\rm rms}}

\def \Hrho {H_{\rho}}

\def \BB {\bm B}

\def \curl {{\bm \nabla}\times}

\def \etatz {\eta^0_{\rm t}}

\def \Rm  {\mbox{Re}_{\rm M}}
\def \Pm  {\mbox{Pr}_{\rm M}}
\def \cs  {c_{\rm s}}
\def \kf  {k_{\rm f}}

\def \urms {u_{\rm rms}}

\def \BB {\bm B}

\def\onethird{{\textstyle{1\over3}}}

\def \ME {E_{\rm M}}

\def \Mh {E^{\rm h}_{\rm M}}
\def \HKh {H^{\rm h}_{\rm K}}
\def \half {{\textstyle{1\over2}}}
\newcommand{\fig}[1]{Fig.~\ref{#1}}
\newcommand{\subfig}[2]{Fig.~\ref{#1}(#2)}

\def\drawing #1 #2 #3 {
\begin{center}
\setlength{\unitlength}{1mm}
\begin{picture}(#1,#2)(0,0)
\put(0,0){\framebox(#1,#2){#3}}
\end{picture}
\end{center} }

\def \apj {ApJ}
\def \apjl {ApJL}
\def \mnras {MNRAS}
\def \pre {Phys. Rev. E}

\def \AandA {A\&A}
\def \jetp {Sov. Phys. JETP}
\def \solphys {Sol. Phys.}

\newtoggle{psfig}
\togglefalse{psfig}

\title[Intense bipolar structures from dynamos]
{Intense bipolar structures from stratified helical dynamos}
\author[D. Mitra, et al.]{
Dhrubaditya Mitra$^{1}$\thanks{E-mail:
dhruba.mitra@gmail.com}, A. Brandenburg$^{1,2}$\thanks{E-mail:
brandenb@nordita.org},
N. Kleeorin $^{3,1,4}$\thanks{E-mail:
nat@bgu.ac.il},
and
I. Rogachevskii $^{3,1,4}$\thanks{E-mail:
gary@bgu.ac.il}
\\
$^1$Nordita, KTH Royal Institute of Technology and Stockholm University,
Roslagstullsbacken 23, SE-10691 Stockholm, Sweden\\
$^2$Department of Astronomy, AlbaNova University Center,
Stockholm University, SE-10691 Stockholm, Sweden\\
$^3$Department of Mechanical Engineering, Ben-Gurion University of the Negev,
POB 653, Beer-Sheva 84105, Israel\\
$^4$Department of Radio Physics, N.~I.~Lobachevsky State University of
Nizhny Novgorod, Russia\\
}
\begin{document}
\date{\today,~ $ $Revision: 1.174 $ $}
%
\maketitle
\label{firstpage}
\begin{abstract}
We perform direct numerical simulations of the equations of magnetohydrodynamics with
external random forcing and in the presence of gravity.
The domain is divided into two parts: a lower layer where the forcing is
helical and an upper layer where the helicity of the forcing is zero
with a smooth transition in between.
At early times, a large-scale helical dynamo develops in the bottom layer.
At later times the dynamo saturates, but the vertical magnetic field
continues to develop and rises to form dynamic
bipolar structures at the top, which later disappear and reappear.
Some of the structures look similar to $\delta$ spots observed in the Sun.
This is the first example of magnetic flux concentrations, owing
to strong density stratification, from
self-consistent dynamo simulations that generate bipolar,
super-equipartition strength, magnetic structures whose energy density
can exceeds the turbulent kinetic energy by even a factor of ten.
\end{abstract}

\section{Introduction}

The most striking and also the most observed
magnetic features of the Sun are the sunspots and active regions.
The number of sunspots, the strength of the magnetic field in
sunspots, and the magnetic field calculated at the
surface of the Sun are often taken as proxies of the solar magnetic
field deep inside.
There is general agreement that the
evolution of the solar magnetic field
is governed by the solar dynamo which operates in the convection zone of the Sun.
This brings us to the question, how is the magnetic field generated by the
solar dynamo related to the magnetic field observed at the surface of the Sun?
At present, this question does not have a clear answer.

The conventional picture \cite[see, e.g.,][for a review]{cho08}
is that the solar dynamo generates a strong toroidal
magnetic field in the form of flux tubes at the bottom of
the convection zone, also called the tachocline.
This strong magnetic field is buoyant and hence rises up to eventually
penetrate through the surface layers of the Sun to create bipolar regions
at the surface.
During its rise through the convection zone,
the magnetic flux tube is twisted by the Coriolis force to give rise to
a preferential tilt of the bipolar regions with respect to the equator
-- which is also known as Joy's law.

The traditional picture is prone to criticism on several counts.
(a) Recent numerical simulations of rotating
spherical magneto-convection \citep{ghi+cha+smo10,kap+man+bra12,aug+bru+mie+too13}
have shown that a solar-like dynamo can operate
in the bulk of the convection zone,
even without a tachocline.
(b) Is it possible for a magnetic flux tube to rise coherently
through the turbulent convection zone and still remain anchored to the
tachocline?
Numerical simulations of \cite{gue+kap11},
admittedly at moderate magnetic Reynolds numbers,
have found no evidence that this is possible.
Recent simulations by \cite{NM14} and \cite{FF14} do find flux loops rising
from mid depths of the convection zone, but in contrast to the traditional
picture, they are not anchored at the bottom of the convection zone\footnote{
\cite{FF14} only show extended patches of toroidal field, so the connection with
sunspot formation remains open.}.
(c) As the flux tube rises, the magnetic field weakens, so even the
traditional picture must invoke a re-amplification
process near the surface.
For example, \cite{par79} postulated downdrafts ``to operate beneath the
sunspot to account for the gathering of flux to form a sunspot.''
Furthermore, current flux emergence simulations that include a photosphere
\citep[see, e.g.,][]{kit+kos+wra+man10, che+rem+tit+sch10, ste+nor12,
rem+che14} do show such re-amplification, but the mechanism
responsible for the re-amplification process remains unknown.
(d) A natural corollary of the rising flux tube picture is that the active
regions will emerge with preferential orientation at the surface of the Sun,
whereas recent observational analysis \citep{ste+kos12} shows that active
regions actually emerge with random orientations but get preferentially
oriented as time progresses.
Note nevertheless that \cite{LC02} have attempted to explain this discrepency 
within the framework of the conventional scenario by arguing that departures
from a preferred orientation are due to turbulent convection and are 
restored past the emergence.  

In the last decade, an alternative scenario has emerged.
In this scenario, first suggested by~\cite{B05}, the turbulent dynamo generates
magnetic field in the bulk of the convection zone.
In the near-surface shear layer, that has been observed in helioseismology \citep{Schou98},
the dynamo-generated magnetic field propagates equatorward, satisfying the
Parker-Yoshimura rule~\citep{par55,yos75}.
The observed preferential orientation of the active regions, the Joy's law,
can be understood as an effect of the shear~\citep{B05}.
In this scenario, which admittedly is yet to be supported by direct numerical
simulations, although mean-field calculations do provide support~\citep{pip+kos11},
the active regions must form from a dynamo-generated large-scale magnetic field by
the process of magnetic flux concentration operating at or near the surface of the Sun.
This process may be the same re-amplification process necessary in the conventional
scenario.

There have been two different, mutually complimentary, approaches to
understand this process.
On the one hand lies the numerical simulations by \cite{kit+kos+wra+man10},
\cite{che+rem+tit+sch10}, \cite{ste+nor12}, and \cite{rem+che14}
who solve radiative magneto-convection in a Cartesian domain under a
simplified setup (non-rotating, no large-scale shear).
All these simulations develop a bipolar magnetic structure at the
top surface, but in all the cases the velocity and the magnetic field at the bottom boundary
need to be carefully imposed.
Furthermore, in these simulations, with the exception of \cite{kit+kos+wra+man10},
the mechanism responsible for formation of magnetic structures
has not been elucidated.
Another related example are the magneto-convection simulations of \cite{tao+wei+bro+pro98},
where an imposed vertical field segregates into magnetized and unmagnetized regions.
The authors ascribe this to the effect of flux expulsion, but the actual mechanism
might well be another one.
On the other hand lies a volume of work \citep{kli+rog+ruz89,
kle+rog+ruz90, kle+rog94,rog+kle07,bra+kle+rog10,bra+kem+kle+mit+rog11, bra+kem+kle+rog12,
bra+kle+rog13, kem+bra+kle+rog+mit12, kem+bra+kle+rog12, kap+bra+kle+man+rog12, war+los+bra+kle+rog13},
which have investigated the possibility  that  the negative effective magnetic pressure
instability (NEMPI) is a mechanism of flux concentration
and formation of active regions.
In all of them, a small (compared to equipartition) background magnetic field
has been imposed in a statistically stationary turbulent magneto-fluid in the presence
of gravity; a large-scale instability (namely NEMPI) develops which forms magnetic
structures.

The essence of this mechanism is related to a negative contribution of turbulence
to the effective magnetic pressure (the sum of non-turbulent and turbulent contributions).
This is caused by a suppression of total (kinetic plus magnetic) turbulent pressure
by the large-scale magnetic field.
For large magnetic and fluid Reynolds numbers these turbulent contributions are large
enough so that the effective magnetic pressure becomes negative.
This results in the excitation of a large-scale instability, i.e., NEMPI.
The instability is efficient if the background magnetic field is within a
specific range, which depends on the relative orientation between gravity and the imposed
field.
The maximum flux concentration achievable depends on the nonlinear saturation of NEMPI;
unipolar spot-like structures \citep{bra+kem+kle+mit+rog11, bra+kle+rog13} and
bipolar active region-like structures \citep{war+los+bra+kle+rog13} have been
obtained under different circumstances.
We emphasize that turbulence plays a crucial role in the formation of those unipolar and
bipolar magnetic structures.
This may seem somewhat counterintuitive because in many other cases turbulence
increases mixing by enhancing diffusion.
However, there is no contradiction because
there are many examples of pattern formation in reaction--diffusion systems
that have been long studied and well understood; see, e.g., \cite{cro+hoh93} for a review.

A shortcoming, that is common between the NEMPI papers and the radiative
magneto-convection papers quoted above
is that the magnetic field is imposed externally, either over the whole volume or
at the lower boundary.
It is then necessary to investigate how the magnetic flux from dynamo-generated
magnetic fields can be
concentrated to form active regions.
Furthermore, it has been observed that NEMPI is suppressed in the presence
of rotation \citep{los+bra+kel+mit+rog12,los+bra+kel+rog12}, which
is an essential ingredient, together with gravity, to the generation of
a large-scale magnetic field by dynamo action.

Hence, it is crucial to study the interaction between NEMPI and large-scale
dynamo instabilities.
It turns out that there exists a range of parameters over which it is possible
for NEMPI to create magnetic flux
concentrations from a dynamo-generated magnetic field; evidence in support of this picture
has been obtained from both mean-field models~\citep{jab+bra+kle+mit+rog13}
and direct numerical
simulations~\citep{jab+bra+los+kel+rog14}.
Particularly interesting cases of flux concentration from dynamo-generated fields,
which have not been studied so far,
are those where dynamo and NEMPI do not operate at the same physical
location, but in different parts of the domain.
For example, the dynamo may operate in the deeper layers of a stratified domain but not
in the upper layers, whereas in the upper layers NEMPI can operate to produce flux
concentrations.
In this paper, we study this problem by direct numerical simulations.

\section{The model}
\label{DNSmodel}

\subsection{Governing equations}

We solve the equations of isothermal magnetohydrodynamics (MHD) for the velocity $\UU$,
the magnetic vector potential $\AAA$, and the density $\rho$,
\begin{equation}
\rho\Dt\UU=\JJ\times\BB-\cs^2\grad\rho+\grad\cdot(2\nu\rho\SSSS)
+\rho(\ff+\grav),
\end{equation}
\begin{equation}
\delt\AAA=\UU\times\BB+\eta\nabla^2\AAA,
\end{equation}
\begin{equation}
\delt\rho=-\grad\cdot\rho\UU,
\end{equation}
where the operator $\Dt \equiv \delt+\UU\cdot\grad$ denotes
the convective derivative,
$\BB=\curl\AAA$ is the magnetic field,
$\JJ=\curl\BB/\mu_0$ the current density,
${\sf S}_{ij}=\half(U_{i,j}+U_{j,i})-\onethird\delta_{ij}\grad\cdot\UU$
is the traceless rate of strain tensor (the commas denote
partial differentiation),
$\nu$ the kinematic viscosity,
$\eta$ the magnetic diffusivity,
and $\cs$ the isothermal sound speed.
In addition, we assume the ideal gas law to hold.
Our domain is a Cartesian box of size $L_x\times L_y\times L_z$
with $L_x=L_y=L_z=2\pi$.
Periodic boundary conditions on all dynamical variables are assumed in the horizontal
($xy$) plane.
The velocity satisfies stress-free, non-penetrating boundary condition
at the top and bottom boundaries.
The volume-averaged density is therefore constant in time and equal to
its initial value.
At the bottom, perfectly conducting boundary conditions are imposed
on the magnetic field, which is constrained to have only a vertical component
at the top boundary (normal field boundary condition).
The gravitational acceleration $\grav=(0,0,-g)$ is chosen such that
$k_1 \Hrho=1$, which leads to a density contrast in the vertical direction
between bottom and top of $\exp(2\pi)\approx535$.
Here $\Hrho\equiv\cs^2/g$ is the density scale height.

\subsection{Forced turbulence}

Turbulence is sustained in the medium by injecting energy through
the function $\ff$ given by \citep{B01}
\begin{equation}
\ff(\xx,t)={\rm Re}\{N\fh(\kk,t)\exp[\ii\kk\cdot\xx+\ii\phi]\},
\end{equation}
where $\xx$ is the position vector.
On dimensional grounds, we choose
$N=f_0 \sqrt{\cs^3 |\kk|}$, where $f_0$ is a
nondimensional forcing amplitude.
At each timestep we select randomly the phase
$-\pi<\phi\le\pi$ and the wavevector $\kk$
from many possible wavevectors
in a certain range around a given forcing wavenumber, $\kf$.
Hence $\ff(t)$ is a stochastic process that is white-in-time and
is integrated by using the Euler--Marayuma scheme \citep{hig01}.
The Fourier amplitudes,
\begin{equation}
\fh({\kk})=\RRRR\cdot\fh({\kk})^{\rm(nohel)}\quad\mbox{with}\quad
{\sf R}_{ij}={\delta_{ij}-\ii\sigma\epsilon_{ijk}\hat{k}
\over\sqrt{1+\sigma^2}},
\end{equation}
where $\sigma$ characterizes the fractional helicity of $\ff$, and
\begin{equation}
\fh({\kk})^{\rm(nohel)}=
\left(\kk\times\eee\right)/\sqrt{\kk^2-(\kk\cdot\eee)^2},
\label{nohel_forcing}
\end{equation}
is a non-helical forcing function, and $\eee$ is an arbitrary unit vector
not aligned with $\kk$ and $\hat{\kk}$ is the unit vector along $\kk$; note that $|\fh|^2=1$.
By virtue of the helical nature of $\ff$, a dynamo develops in the
domain~\citep{B01}.
As we want to separate the domain over which dynamo operates from the
domain over which it is possible for magnetic flux concentrations to happen, we
choose the fractional helicity of the force $\sigma$ to go to zero
at the top layers of our domain, i.e., for $z > z_0$,
v.i.z.,
\begin{equation}
\sigma(z-z_0) = \frac{\smax}{2}\left[1-\erf\left(\frac{z-z_0}{\wf}\right) \right].
\end{equation}
Here $\erf$ is the error function, and $\wf$ is a length scale chosen to be
$0.08L_z$.
We use several different values of $z_0$ and $\smax$.

\subsection{Non-dimensional parameters}

We choose our units such that $\mu_0=1$ and
$\cs=1$.
Our simulations are characterized by the
fluid Reynolds number $\Rey\equiv\urms/\nu\kf$,
the magnetic Prandtl number $\Pm=\nu/\eta$, and
the magnetic Reynolds number $\Rm\equiv \Rey \, \Pm$.
The magnetic field is expressed in units of
$\Beqz\equiv\sqrt{\rho_0} \, \urms$.
As the value of the turbulent velocity is set by the local strength
of the forcing, which is uniform, the turbulent velocity is also
statistically uniform over depth, and therefore we choose to define
$\urms$ as the root-mean-square velocity based on a volume average
in the statistically steady state.
On the other hand, the density varies over several orders of magnitude as
a function of depth and hence we choose $\rho_0$ as the horizontally and
temporally average density at $z=0$, which is the middle of the domain.
Time is expressed in eddy turnover times, $\Tto=(\urms\kf)^{-1}$.
We often find it useful to consider the turbulent-diffusive timescale,
$\Teta=(\etatz k_1^2)^{-1}$, where $\etatz=\urms/3\kf$ is
the estimated turbulent magnetic diffusivity.

The simulations are performed with the {\sc Pencil Code},%
\footnote{{\tt http://pencil-code.googlecode.com}}
which uses sixth-order explicit finite differences in space and a
third-order accurate time stepping method.
We typically use a numerical resolutions of $256^3$ mesh points,
although some representative simulations at higher resolutions are
also run.

\begin{table}\caption{
Summary of the runs discussed in the paper.
Here, $\tilde\lambda=\lambda/\urms\kf$ is a nondimensional growth rate.
}\vspace{12pt}\centerline{\begin{tabular}{lrcrclcr}
Run & $z_0$ & $\smax$ & $\Rm\!\!$ & $\tilde\kf$ & $~~\tilde\lambda$ & $\Tto$ & $\Teta$  \\
\hline
A    &  2 &   1  & 17 & 30 & 0.041 & 0.33 &  900 \\
B    &$-1$&   1  & 17 & 30 & 0.042 & 0.33 &  900 \\
B/2  &$-1$&   1  & 17 & 30 & 0.036 & 0.33 &  900 \\
C    &$-2$&   1  & 17 & 30 & 0.045 & 0.33 &  900 \\
D    &$-2$&   1  & 17 & 60 & 0.043 & 0.17 & 1800 \\
E    &$-2$&   1  & 170& 30 & 0.022 & 0.33 &  900 \\
0-02 & 0  & 0.2  & 17 & 30 & 0.0043& 0.33 &  900 \\
0-1  & 0  &  1   & 17 & 30 & 0.043 & 0.33 &  900 \\
\label{Ttimescale}\end{tabular}}\end{table}

\section{Results}

\begin{figure}
\iftoggle{psfig}{
  \includegraphics[width=0.95\columnwidth]{fig/runB_ME.ps}
  \put(-50,120){\Large{(a)}} \\
  \includegraphics[width=0.95\columnwidth]{fig/runB_ME_inset.ps}
  \put(-60,150){\Large{(b)}}
 }{
  \includegraphics[width=0.95\columnwidth]{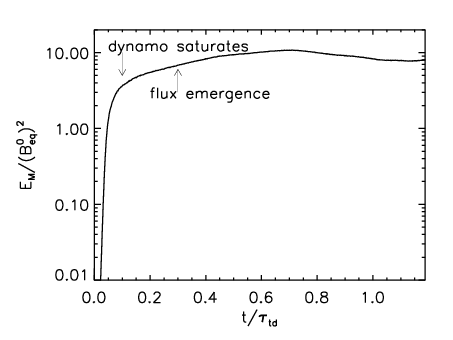}
  \put(-50,120){\Large{(a)}} \\
  \includegraphics[width=0.95\columnwidth]{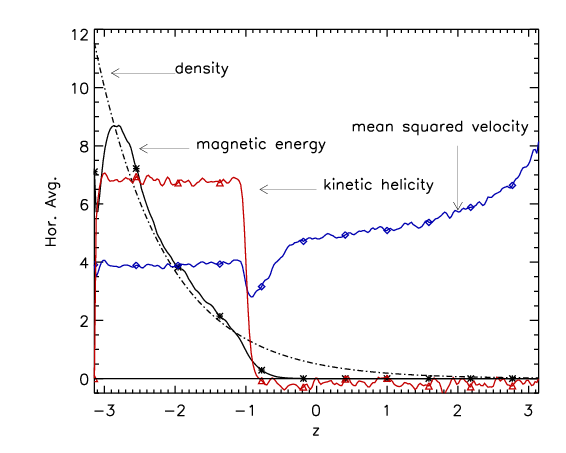}
  \put(-60,150){\Large{(b)}}
}
\caption{(Color online) (a) Evolution of magnetic energy, $\ME$ from {\tt Run~B}.
(b) Non-dimensional values of horizontally averaged (averaged
over the $xy$ plane) density (broken line), $\brah{\rho}/\brah{\rho(z=0)}$,
mean squared velocity (blue $\diamond$), $\brah{{\bm U}^2}/\cs^2$,
magnetic energy ($\ast$), $\Mh/(\Beqz)^2$, and kinetic helicity (red, $\triangle$)
$\HKh/\kf\urms^2$ as a function of the
height $z$ at dynamo saturation, i.e., at $t/\Teta=0.1$ from {\tt Run~B}.
For clarity, the density, the mean squared velocity, and the kinetic helicity are
scaled by a factor of  $1/2$, $600$ and $10$ respectively.
}\label{fig:ME}
\end{figure}

We have performed a number of runs varying mainly the values of
$z_0$ and $\sigma$.
We always used $\Pm=0.5$ and, in most of the cases, we had
$\Rm=17$ and $\tilde\kf\equiv\kf/k_1=30$, but in one case
we also used $\Rm=170$ and in another $\tilde\kf=60$.
Our runs are summarized in Table~\ref{Ttimescale}.
Let us start by describing in detail one representative simulation
among the many we have run; v.i.z., the case of {\tt Run~B} in
Table~\ref{Ttimescale}.
In this case, the flow is {\it helically} forced up to the height of $z_0/H_\rho = -1$
with $\smax=1$.
Above the plane $z=z_0$ the flow is indeed forced, but not helically,
i.e., with $\sigma=0$.
By virtue of helical forcing from the bottom wall up to the height
of $z_0$, a dynamo develops.
In \fig{fig:ME} we show the evolution of the volume averaged magnetic energy, $\ME$,
defined by
\begin{equation}
\ME = \frac{1}{V} \int_V d {\bm r} \, \half B^2.
\label{eq:KM}
\end{equation}
At short times there is a fast exponential growth of $\ME$;
the growth rate, $\lambda$, is given in Table~\ref{Ttimescale}.
The dynamo saturates at about $0.1 \Teta$, see \fig{fig:ME}(a).
In \fig{fig:ME}(b), we show the variation of horizontally averaged
(over the $xy$ plane) density $\brah{\rho}$,
mean squared velocity $\brah{{\bm U}^2}$,
magnetic energy $\Mh\equiv \half \brah{{\bm B}^2}$,
and kinetic helicity $\HKh\equiv\brah{\WW\cdot\UU}$
as a function of the height $z$,
where $\WW\equiv\curl\UU$ is the vorticity.
It is clear from \fig{fig:ME}(b) that immediately after dynamo saturation,
both the kinetic helicity and the magnetic field are
largely confined within the domain up to the height $z_0$,
but not the kinetic energy of the turbulence.
Furthermore, in the deep parts of the domain, the horizontally averaged
magnetic energy density is approximately proportional to density and
thus to the local equipartition value,
$\Beq (z) \equiv \brah{\rho{\bm U}^2}^{1/2}$.

\subsection{Flux emergence at the top surface}

As the simulation progresses, at $t/\Teta \approx 0.3$, magnetic flux of
both signs emerges on the top surface.
At first the flux emerges as small-scale fluctuations, but within a time of
about $0.1\Teta$, it self-organizes to a bipolar structure.
The two polarities of the bipolar structure then move away from each other.
This is demonstrated in a series of snapshots shown in \fig{fig:emerg}.
Here, stratified turbulence gives rise to anti-diffusive properties
leading to the formation of bipolar structures.
This is the first remarkable result from our simulations.
Similar behaviour has been seen by \cite{ste+nor12}, although
not in self-consistent dynamo simulations but in simulations
where the magnetic field at the bottom boundary
was imposed in the upwellings.
Furthermore, the self-organization we observe is not driven by
radiative convection, as in the simulations of \cite{ste+nor12}
but by forced isothermal turbulent flows.

\subsection{Formation of an intense bipolar structure}

\begin{figure*}
\iftoggle{psfig}{
  \includegraphics[width=0.5\columnwidth]{fig/runB1_emerge_0270.ps}
  \includegraphics[width=0.5\columnwidth]{fig/runB1_emerge_0280.ps}
  \includegraphics[width=0.5\columnwidth]{fig/runB1_emerge_0290.ps}
  \includegraphics[width=0.5\columnwidth]{fig/runB1_emerge_0300.ps}
 }{
\includegraphics[width=0.5\columnwidth]{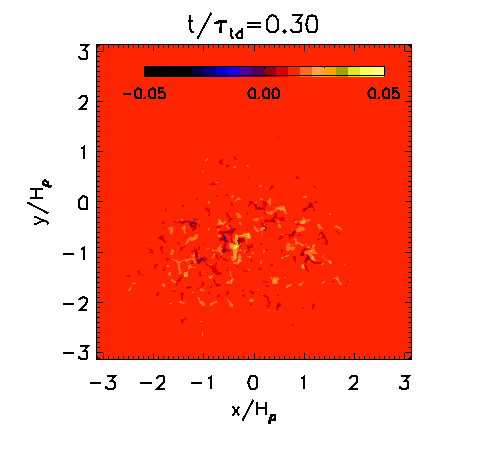}
\includegraphics[width=0.5\columnwidth]{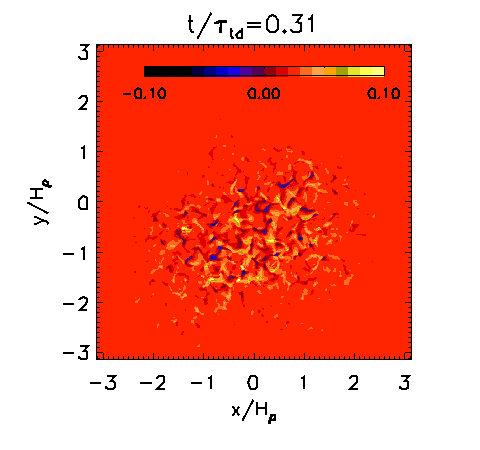}
\includegraphics[width=0.5\columnwidth]{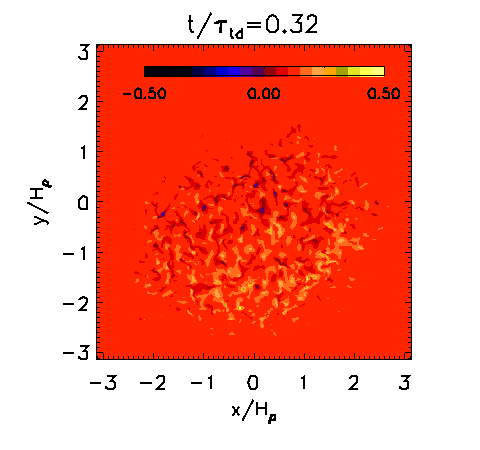}
\includegraphics[width=0.5\columnwidth]{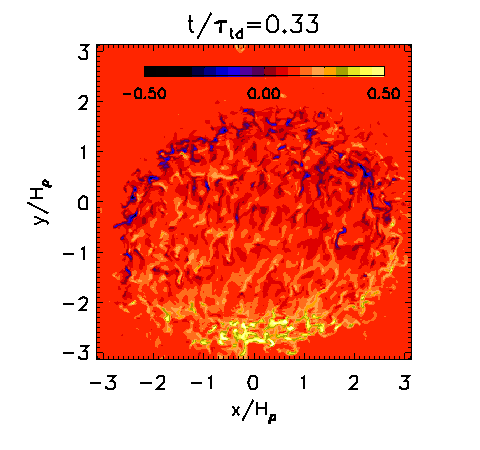}
}
\caption{Vertical magnetic field at the top surface
at different times (from $t/\Teta = 0.30$ to $0.33$) from {\tt Run~B}.
The magnetic field is normalized by $\Beqz$. }
\label{fig:emerg}
\end{figure*}

Due to periodic boundary conditions in the $x$ and $y$ directions,
the two polarities, while
moving away from each other, approach each other across the far end of the periodic domain,
come close to each other and form a curious bipolar structure,
reminiscent of the so-called $\delta$ spots \citep[see review by][]{fisher}.
The $z$ component of the magnetic field is close to three times $\Beqz$.
This is shown in a series of snapshots in \fig{fig:spot},
where we have shifted the coordinate system relative to the one in \fig{fig:emerg}
so as to have the
bipolar structure in the middle of the top surface.
As we are using periodic boundary conditions along
the horizontal directions, we are free to make such a shift.
To illustrate this, we show in \fig{fig:periodic} the magnetic field at
the top of our computational domain in a box that is extended periodically
to three times its originally size in both the $x$ and $y$ directions.

\subsection{Recurrent spot activity}

\begin{figure*}
\iftoggle{psfig}{
  \includegraphics[width=0.5\columnwidth]{fig/runB1_emerge_0315.ps}
  \includegraphics[width=0.5\columnwidth]{fig/runB1_emerge_0320.ps}
  \includegraphics[width=0.5\columnwidth]{fig/runB1_emerge_0330.ps}
  \includegraphics[width=0.5\columnwidth]{fig/runB1_emerge_0340.ps}
 }{
\includegraphics[width=0.5\columnwidth]{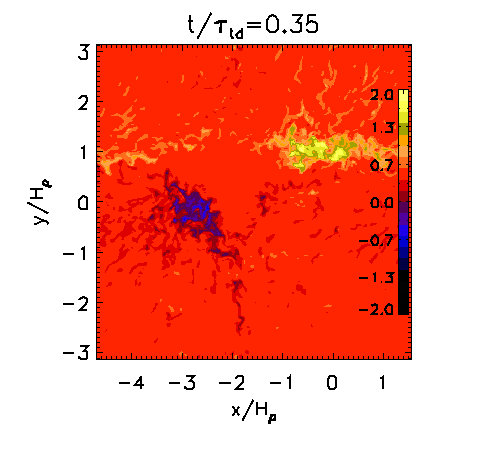}
\includegraphics[width=0.5\columnwidth]{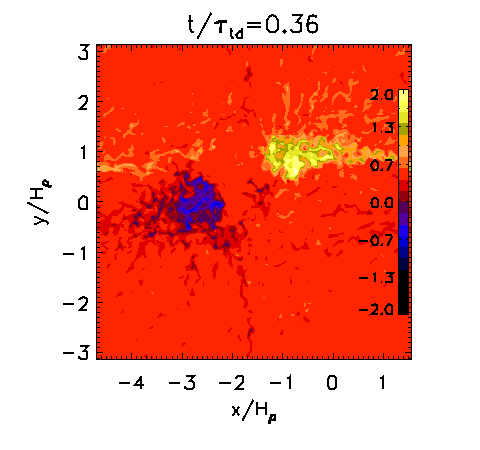}
\includegraphics[width=0.5\columnwidth]{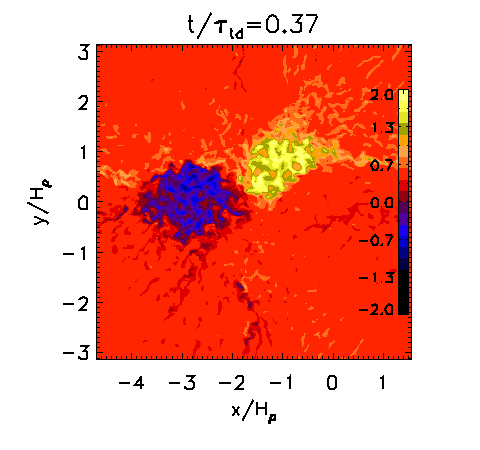}
\includegraphics[width=0.5\columnwidth]{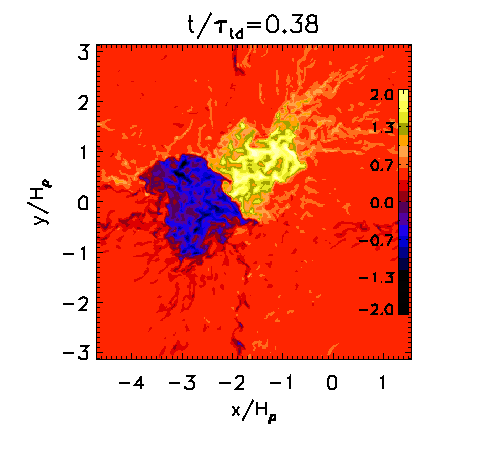}
}
\caption{Same as \fig{fig:emerg}, but at later times
(from $t/\Teta = 0.35$ to $0.38$) and the frame is re-centered,
as illustrated in \fig{fig:periodic} below.
}\label{fig:spot}
\end{figure*}

\begin{figure}
\iftoggle{psfig}{
 \includegraphics[width=\columnwidth]{fig/runB1_emerge_periodic.ps}
}{
\includegraphics[width=\columnwidth]{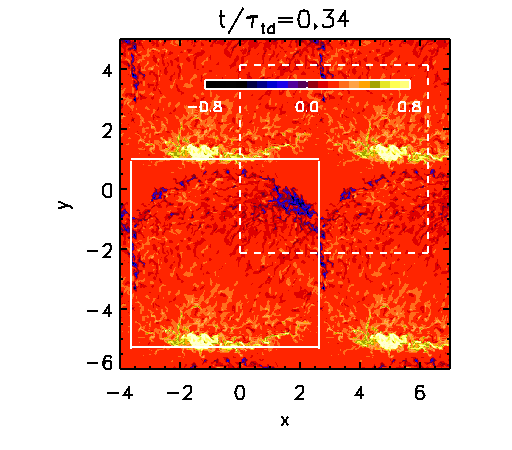}
}
\caption{Vertical magnetic field at the top surface at
$t/\Teta=0.34$ from {\tt Run~B}. The domain has been extended periodically along both
the $x$ and $y$ directions. The solid lines draw the box
used in \fig{fig:emerg} and the dashed lines draw the box used in \fig{fig:spot}.
The magnetic field is normalized by $\Beqz$.
}\label{fig:periodic}
\end{figure}

This spot-like structure survives up to $t/\Teta \approx 0.45$, after which
it turns into a bipolar band whose evolution is shown in  a series of snapshots
in \fig{fig:band}.
At about $t/\Teta\approx1.2$ the band dissolves and the field at the top surface
is close to zero.
And a little while later the band-like structure reappears at a
different position on the top surface and with time evolves to a spot-like
structure similar to the one shown in \fig{fig:spot}; compare the last snapshot
shown in \fig{fig:band} with that of \fig{fig:spot}.

\subsection{How generic are the observed magnetic structures?}

\begin{figure}
\iftoggle{psfig}{
  \includegraphics[width=0.49\columnwidth]{fig/runBby2_surbz0.ps}
  \includegraphics[width=0.49\columnwidth]{fig/runBby2_surbz1.ps}
}{
  \includegraphics[width=0.49\columnwidth]{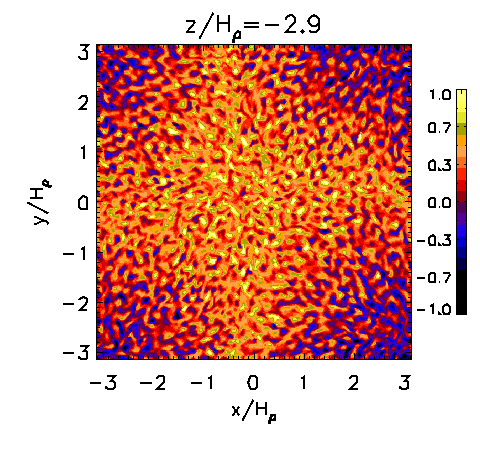}
  \includegraphics[width=0.49\columnwidth]{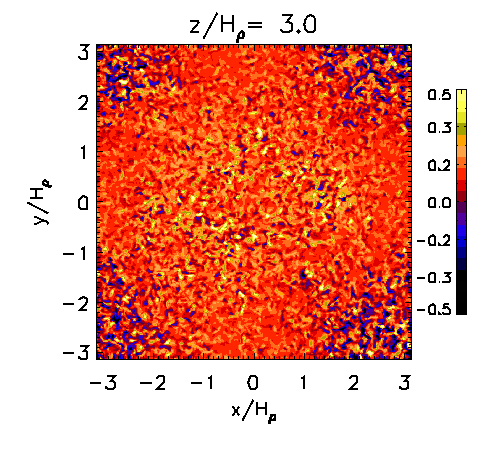}
}
\caption{Contour plot of $B_z/\Beqz$ from {\tt Run~B/2} at two
  different heights.
}\label{fig:runBby2_2surf}
\end{figure}

To summarise, in this simulation, {\tt Run~B},  the normal magnetic field at the
top surface shows  three principal qualitative features:
(a) flux emergence, (b) formation of bipolar structures (spots and bands)
and (c) a recurrent but not exactly periodic appearance of the bipolar structures.
How typical are these qualitative behaviours with respect to variation of various parameters
of our simulation?
This question is addressed in the following manner:
(a) We run a simulation, {\tt Run~O-02}, with the same parameters of {\tt Run~B}
but with a different fractional helicity, $\sigma=0.2$.
For this run,
the helical dynamo instability is excited at a slower rate and the magnetic flux emergence
at the top surface happens at a later time, nevertheless the same qualitative feature
of bipolar magnetic structures are observed.
(b) Keeping the value of fractional helicity, $\smax=1$, to be constant,
we vary the height of the dynamo
region, $z_0/\Hrho$ from $-1$ ({\tt Run~B}) to $-2$ ({\tt Run~C}), $0$ ({\tt Run~O-1}),
and $2$ ({\tt Run~A}).
The flux emergence happens at different times; for higher $z_0$ the flux emergence is faster.
Other than this quantitative change, there is no qualitative change to our results.
(c) We run a simulation {\tt Run~E} with the same parameters as {\tt Run~C},
but with bigger resolution ($384^3$)
and higher Reynolds number and obtain the same qualitative behaviour.
In another simulation, {\tt Run~D}, we keep all the parameters the same
as {\tt Run~E}, except for the forcing wavenumber,
$\tilde\kf=60$, and obtain the same qualitative behaviour.
(d) Finally, we note that {\em gravity plays a crucial role}. In simulations
without gravity ($g=0$) or even $g/\cs^2 k_1=1/2$ ({\tt Run~B/2}),
no sharp magnetic structures are seen.
Instead the magnetic field at the top has the same length scales as the dynamo-generated
magnetic field at the bottom part of the domain,
as demonstrated in \fig{fig:runBby2_2surf}.
It is also clear from our results that the bipolar magnetic structures are strongly
influenced by the periodicity of our domain.
Is it possible to obtain similar structures, but at different length scales
(relative to the box size) and in a larger domain?
By running a simulation with double the box size ($L_x=L_y=L_z=4\pi$) we have found that
the characteristic length scales of the bipolar structures scaled by the box-size remains
the same.
This is because in our periodic geometry, the scale of the large-scale
dynamo is always the largest possible one that fits into the domain.
In future work, it is therefore important to relax this constraint arising
from periodic boundary conditions using, for example, spherical geometry.

\subsection{Sharp bipolar structures}

\begin{figure*}
\iftoggle{psfig}{
  \includegraphics[width=0.32\linewidth]{fig/runB1_emerge_0400.ps}
  \put(-120,120){\Large{(a)}}
  \includegraphics[width=0.33\linewidth]{fig/runB1_emerge_0750.ps}
  \put(-120,120){\Large{(b)}}
  \includegraphics[width=0.33\linewidth]{fig/runB1_emerge_0980.ps}
  \put(-120,120){\Large{(c)}}\\
  \includegraphics[width=0.32\linewidth]{fig/runB1_emerge_1100.ps}
  \put(-120,120){\Large{(d)}}
  \includegraphics[width=0.33\linewidth]{fig/runB1_emerge_1200.ps}
  \put(-120,120){\Large{(e)}}
  \includegraphics[width=0.33\linewidth]{fig/runB1_emerge_1500.ps}
  \put(-120,120){\Large{(f)}}
}{
  \includegraphics[width=0.32\linewidth]{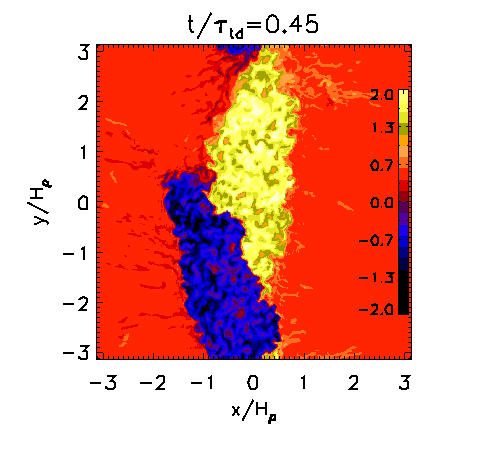}
  \put(-120,120){\Large{(a)}}
  \includegraphics[width=0.33\linewidth]{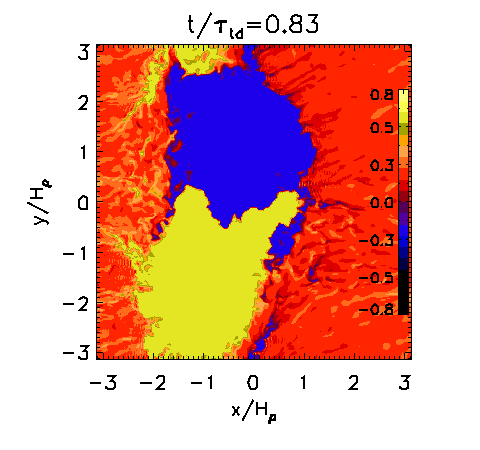}
  \put(-120,120){\Large{(b)}}
  \includegraphics[width=0.33\linewidth]{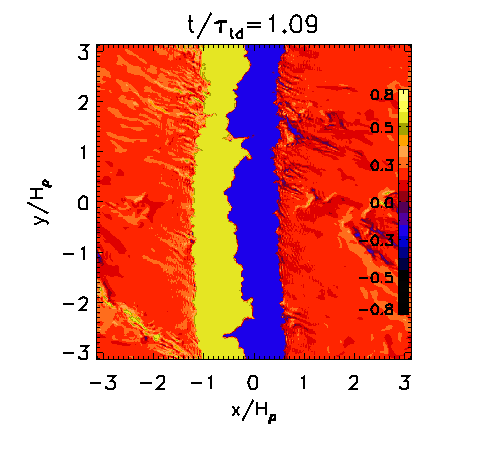}
  \put(-120,120){\Large{(c)}}\\
  \includegraphics[width=0.32\linewidth]{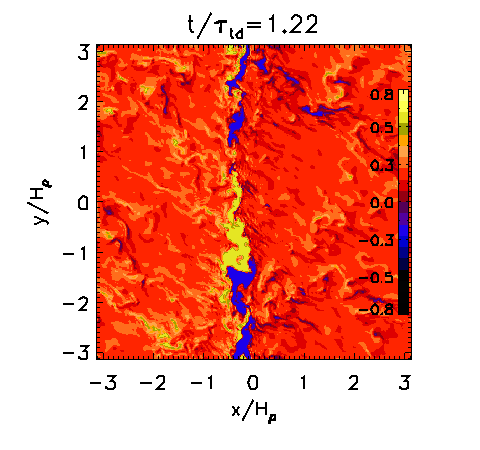}
  \put(-120,120){\Large{(d)}}
  \includegraphics[width=0.33\linewidth]{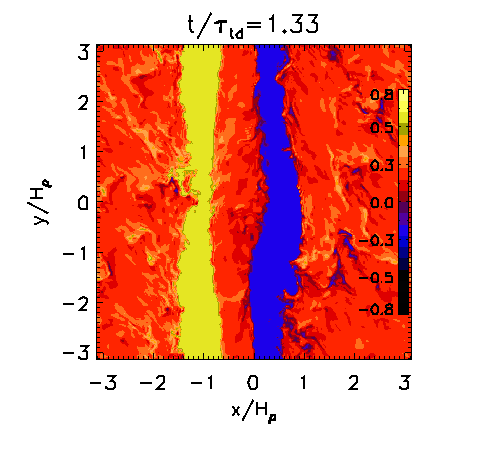}
  \put(-120,120){\Large{(e)}}
  \includegraphics[width=0.33\linewidth]{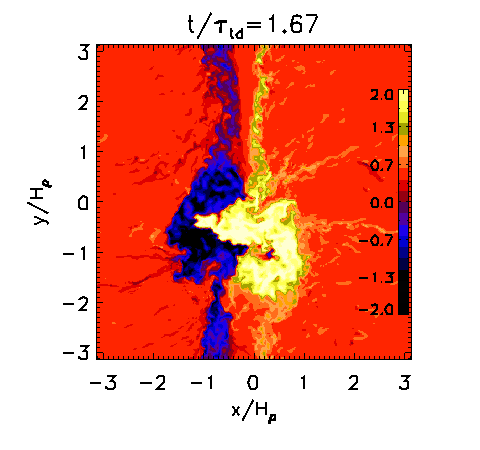}
  \put(-120,120){\Large{(f)}}
}
\\
\caption{Evolution of the vertical magnetic field at the top surface.
Snapshots at different times (from $t/\Teta = 0.45$ to $t/\Teta = 1.67$)
are plotted.  }
\label{fig:band}
\end{figure*}

\begin{figure}
\iftoggle{psfig}{
  \includegraphics[width=\columnwidth]{fig/runB1_angavgspec.ps}
}{
  \includegraphics[width=\columnwidth]{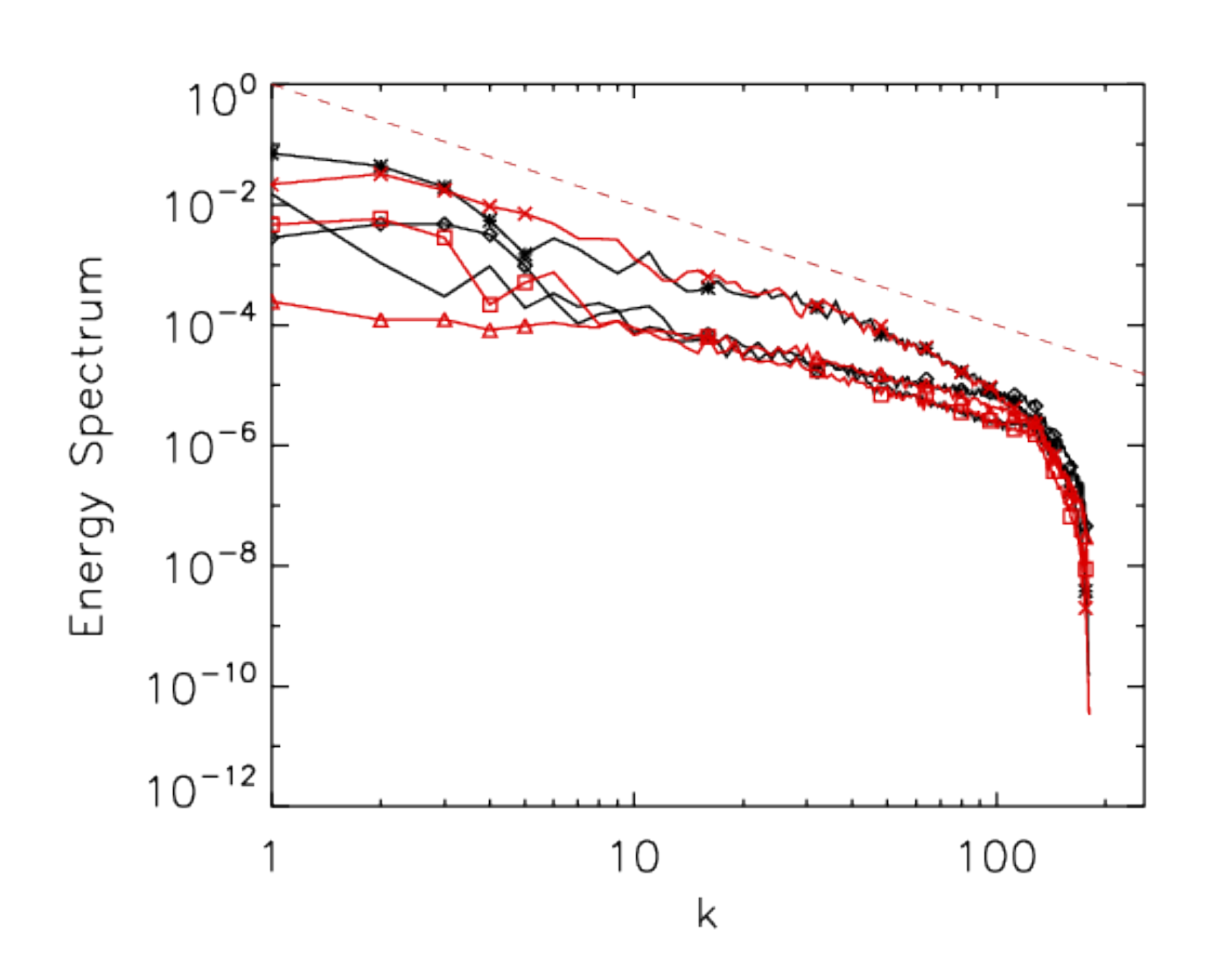}
}
\caption{Angle integrated power spectrum of $B_z$ at the top surface
of our computational box from {\tt Run~B} in log-log scale.
The three black lines show the early times $t/\Teta=0.45$ ($\ast$),
0.83 (no symbol),
and 1.09 ($\diamond$), while the three red lines show the later times
$t/\Teta=1.22$ ($\triangle$), 1.33 ($\square$), 1.67($\times$).
The dashed lines has slope equal to $-2$.
}\label{fig:ang_spec}
\end{figure}

A particularly interesting aspect of these simulations is the formation of bipolar
magnetic structures with sharp edges, examples of which are \fig{fig:spot} or
\fig{fig:band}.
To document the characteristic length scale appearing in magnetic
structures, we plot in \fig{fig:ang_spec} the angle-averaged Fourier spectrum of
$B_z$ at the top surface at different times corresponding to the
snapshots in \fig{fig:band}.
The plot demonstrates that, to
represent the sharp structures, e.g., in the last snapshot in
\fig{fig:band}, Fourier modes up to $k_x/k_1 = 10$ and
$k_y/k_1 = 10$ are necessary.
This also underscores the necessity of having a large scale separation ($\kf/k_1 = 30$)
to see these magnetic structures.
Furthermore, we find that at large $k$, the spectra can be
approximated by a $k^{-2}$ power law.

\begin{figure}
\iftoggle{psfig}{
\includegraphics[width=0.98\columnwidth]{fig/ver2_bb_2surf_flines_dynamo256k30VF_relhel1_roff2.ps}
}{
\includegraphics[width=0.98\columnwidth]{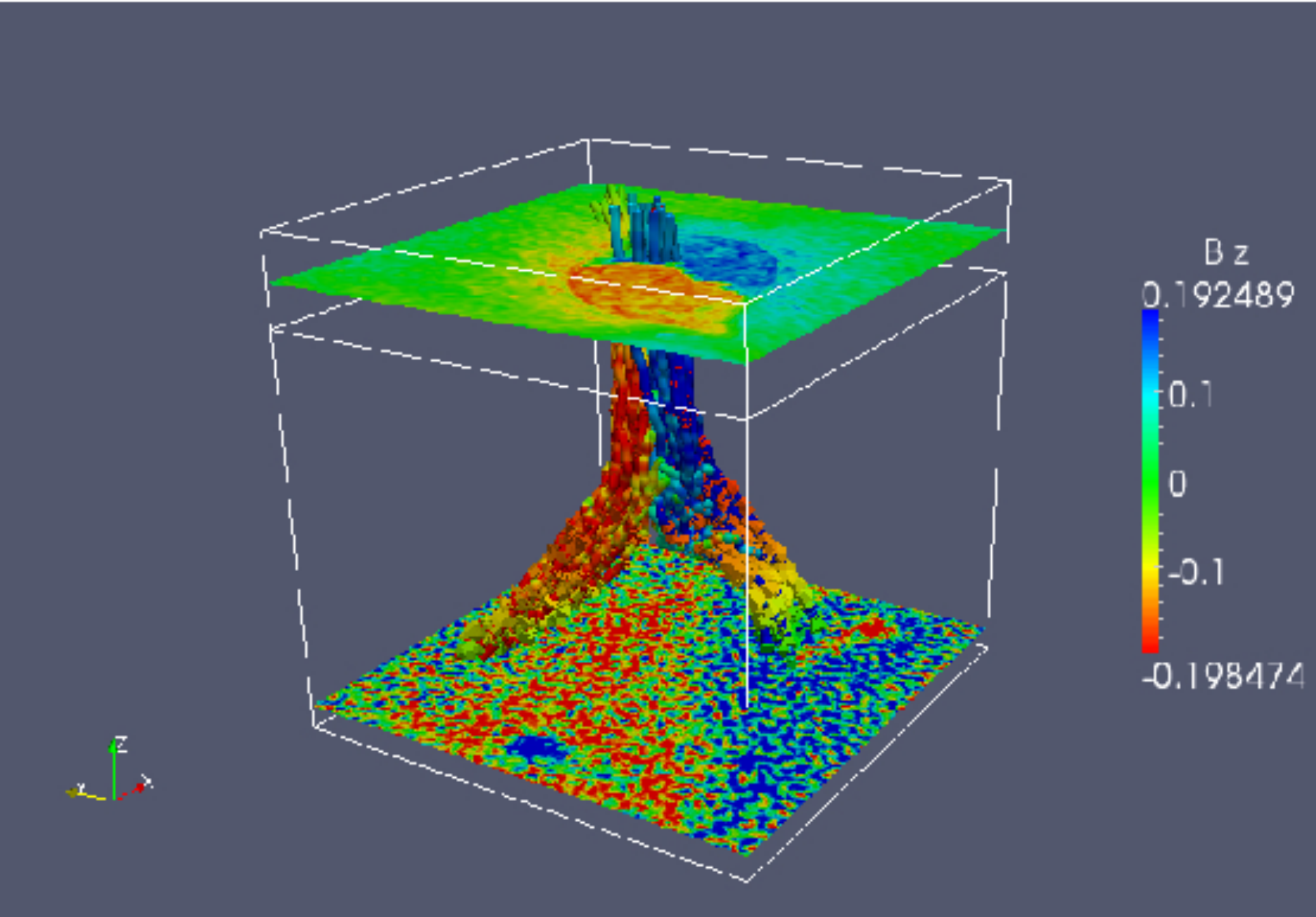}
}
\caption{
Magnetic field structure for {\tt Run~A} at time $t/\Teta \approx 1.2$.
The $z$ component of the magnetic field, $B_z$ is plotted at $z/\Hrho=3.$.
The height up to which dynamo operates,  $z_0/\Hrho=2$, is also shown as a frame.
Here magnetic field, $B_z$ is not normalized, but in units of
$\sqrt{\brah{\rho(z=0)}}\cs$. In the same units $\Beqz\approx0.1$.
}\label{fig:spot3d}
\end{figure}

\begin{figure}
\iftoggle{psfig}{
  \includegraphics[width=.9\columnwidth]{fig/runA_UOmega_top.ps}
}{
  \includegraphics[width=.9\columnwidth]{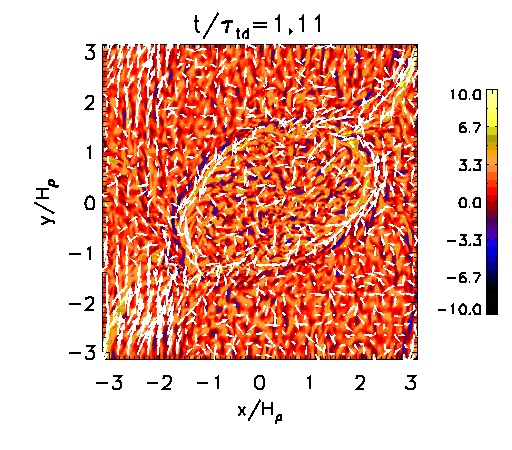}
}
\caption{Contours of the vertical component of vorticity and the
horizontal component of velocity (as arrows)
from {\tt Run~A} at the plane $z/\Hrho=3$; the magnetic structure at the
same plane at the same time, shown in \fig{fig:spot3d}, can be clearly identified.
}\label{fig:spot_vel}
\end{figure}

To take a closer look at the bipolar structure, we show in \fig{fig:spot3d} the spot-like
structure from {\tt Run~A} plotted together with the magnetic field lines in a
three-dimensional representation.
The magnetic field lines of opposite orientation approach each other with
height and merge into a single sharp spot-like structure.
This magnetic structure leaves a clear signature on the velocity field as we demonstrate
in \fig{fig:spot_vel} by plotting the contours of the vertical component of $\WW$
overlaid with the horizontal components of velocity as arrows from {\tt Run~A}.

\subsection{Can NEMPI describe our numerical results?}

Let us now try to understand the flux emergence and the formation of bipolar structure.
This falls in the general class of pattern formation in turbulent systems.
A theoretical technique to describe this general class of problems is the
mean-field theory where we average over the turbulent state to derive a set
of mean-field equations.
The problem of pattern formation then becomes a problem of studying
the instabilities using the mean-field equations.
A well-known example, pioneered by
\cite{kra+rad+ste71} and \cite{Kra+Rad80}
is that of dynamo theory
where the mean-field theory is applied to the induction
equation \cite[see, e.g.,][for a review]{BS05}.
A recent example of an application of this method to understand magneto-rotational instability
in the presence of small-scale turbulence is by \cite{vai+bra+mit+kap+man13}.

For the present problem, we need to average the momentum equation over the statistics
of turbulence.
As a result of such an averaging, a new term
(describing the turbulent contributions) will be added to the large-scale
magnetic pressure term \citep{kle+rog+ruz90, kle+rog94,rog+kle07}.
It has been shown that the effective magnetic pressure that is the sum of non-turbulent
and turbulent (new term) contributions,
can be negative in the presence of a background magnetic field which,
in this problem, will be provided by the dynamo.

\begin{figure}
\iftoggle{psfig}{
  \includegraphics[width=0.9\columnwidth]{fig/runB1_Beq_emergence.ps}
  \put(-160,60){\Large{(a)}} \\
  \includegraphics[width=0.9\columnwidth]{fig/runB1_Beq_late.ps}
  \put(-160,100){\Large{(b)}}
}{
  \includegraphics[width=0.9\columnwidth]{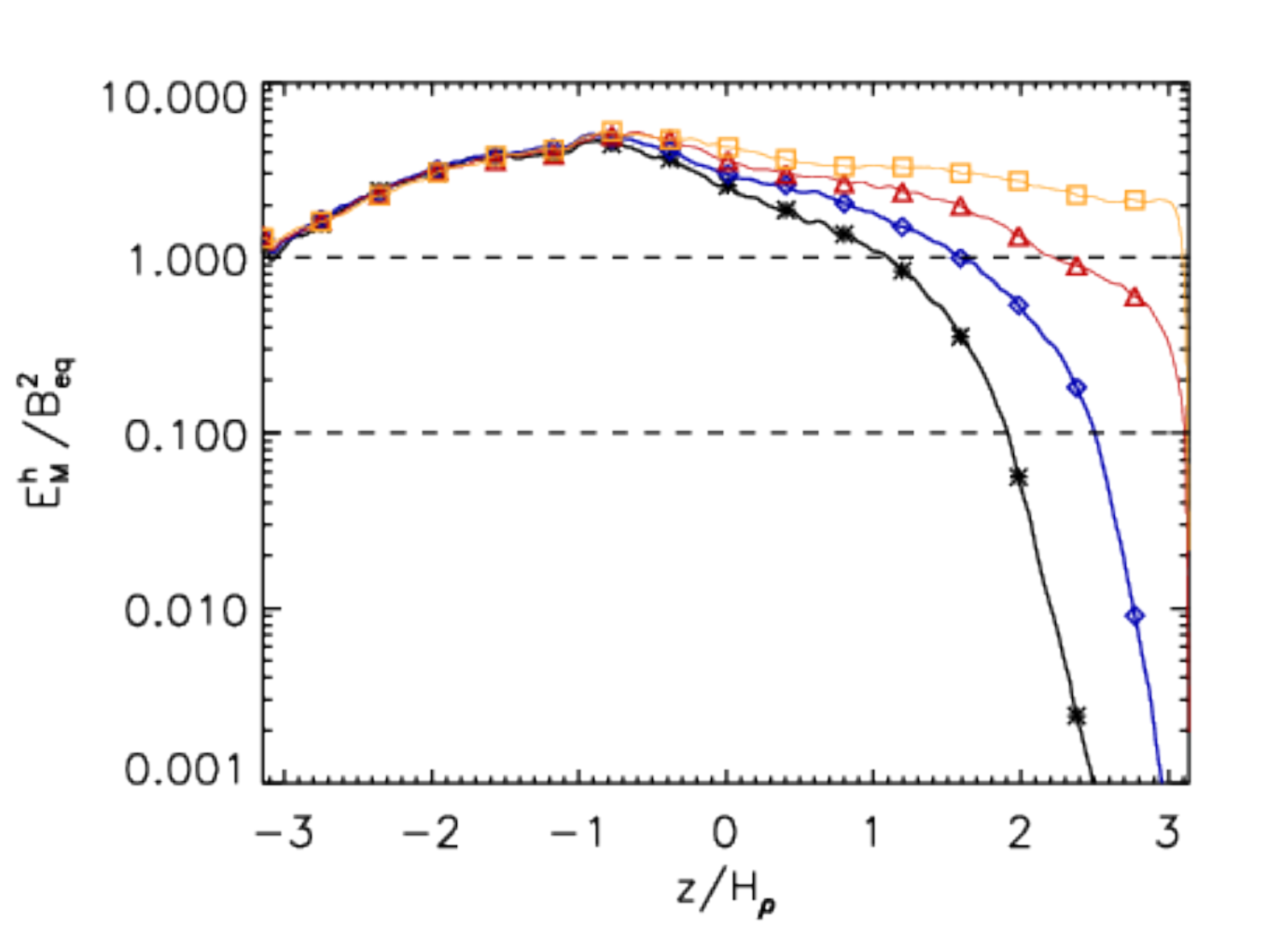}
  \put(-160,60){\Large{(a)}} \\
  \includegraphics[width=0.9\columnwidth]{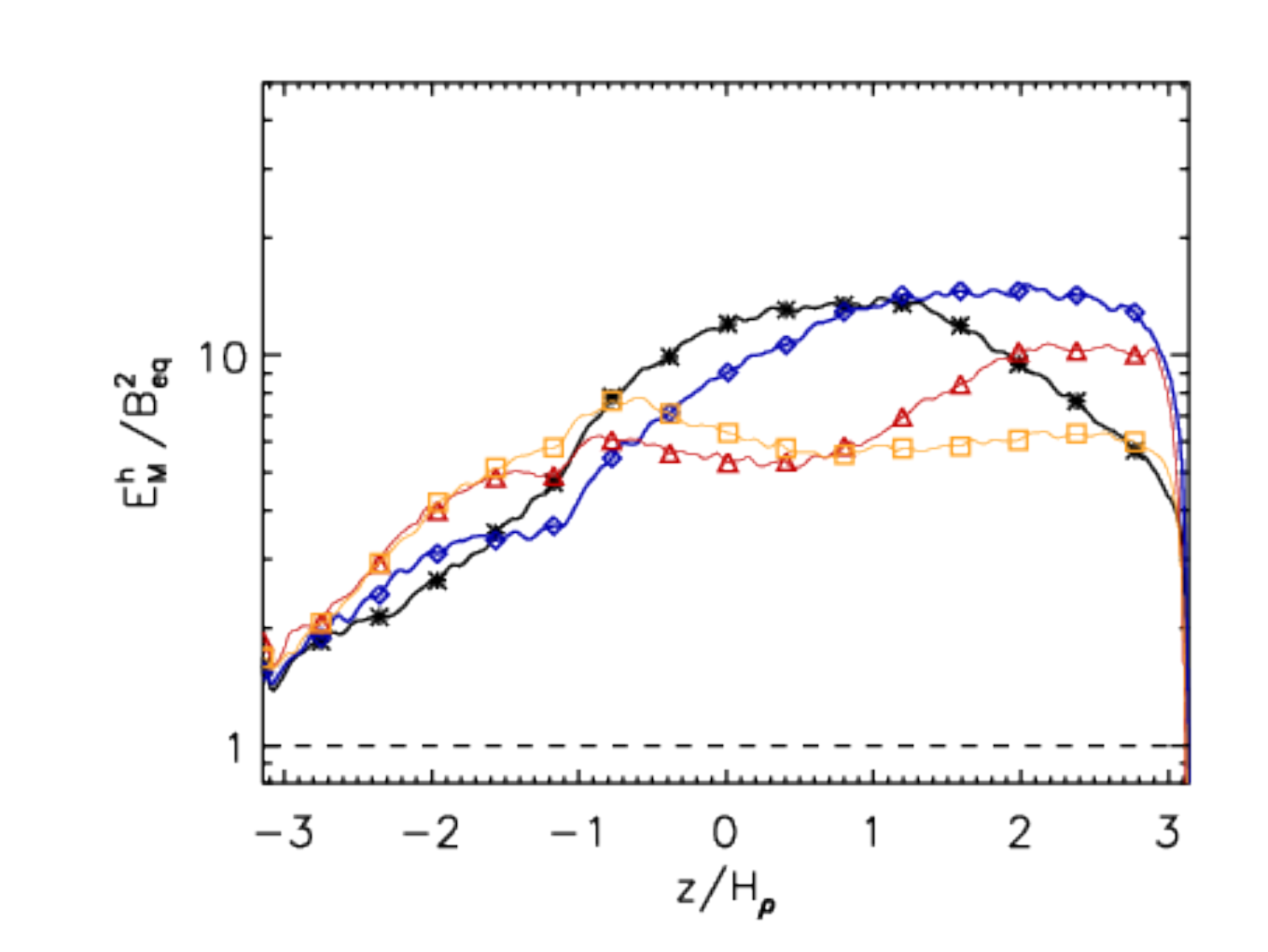}
  \put(-160,100){\Large{(b)}}
}
\caption{ (a) Log-linear plot of horizontally averaged magnetic energy $\Mh(z)$ normalized by the
equipartition value of magnetic energy at height $z$,
$\Beq (z) \equiv\brah{\rho U^2} $,
as a function of the height $z$ at different times
$t/\Teta=0.28 (\ast)$,  $0.30 (\diamond)$,  $0.32 (\triangle)$, and $0.34 (\square)$.
The two dashed lines shows that range of values over which NEMPI can operate effectively.
(b) The same plot, but this time corresponding to the snapshots
plotted in \fig{fig:band};
$t/\Teta=0.86 (\ast)$,  $1. (\diamond)$,  $1.2 (\triangle)$, and $1.33 (\square)$.
}
\label{fig:beq_diff_times}
\end{figure}

\begin{figure}
\iftoggle{psfig}{
  \includegraphics[width=0.98\linewidth]{fig/runB1_bxby_bfly.ps}
}{
  \includegraphics[width=0.98\linewidth]{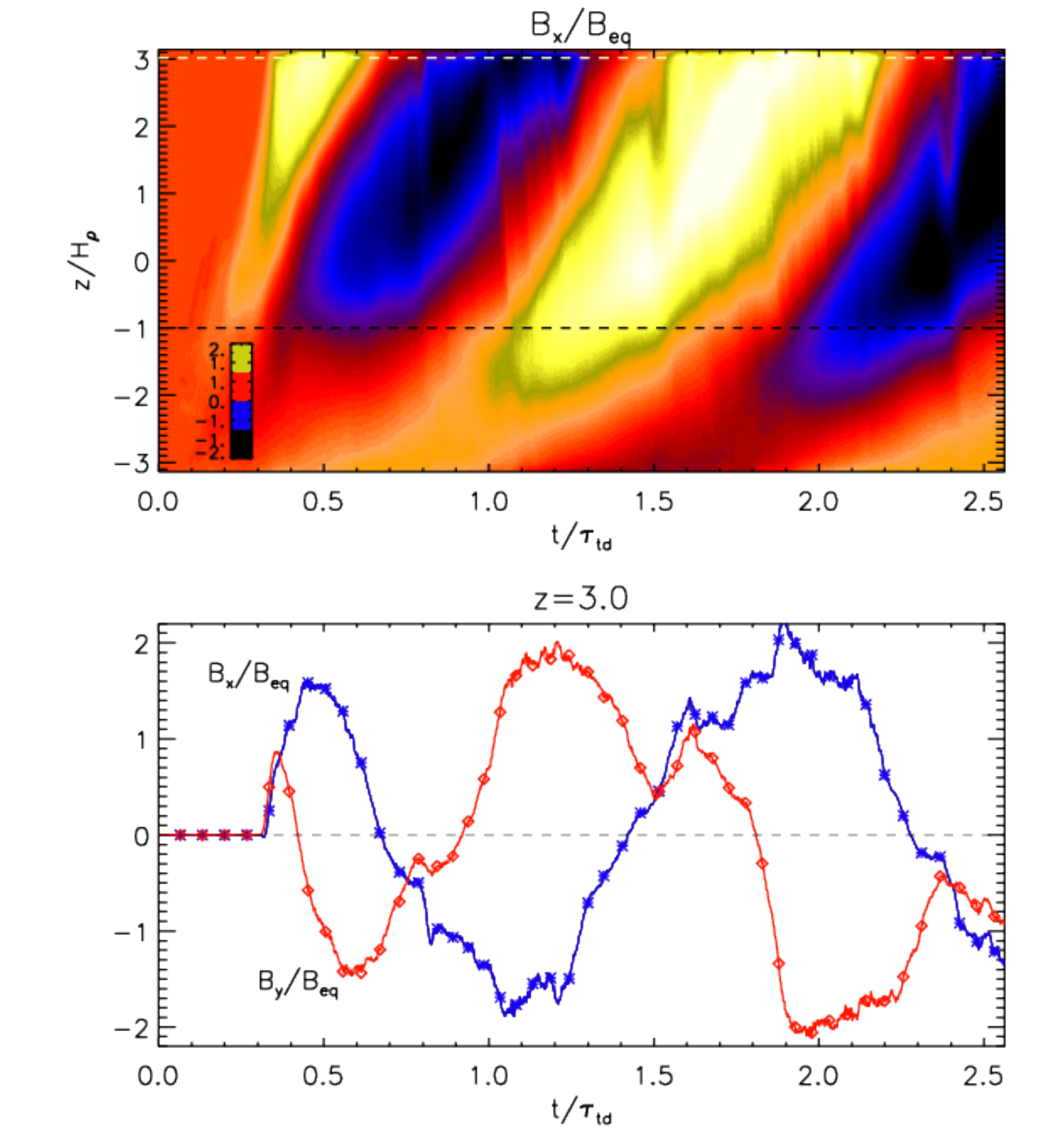}
}
\caption{ (a) Space-time diagram of $\Bhx/\Beq$ showing dynamo waves
  propagating vertically outward. (b) $\Bhx(z)/\Beq(z)$ ($\ast$) and
$\Bhx(z)/\Beq(z)$ ($\diamond$) as a function of time at $z/\Hrho =
3$ .
}
\label{fig:bxby_bfly}
\end{figure}

From symmetry arguments, such a term can be constructed using the background magnetic
field and gravity.
In the two extreme cases: one in which the gravity and the background magnetic field
are perpendicular to each other \citep{bra+kem+kle+rog12, kap+bra+kle+man+rog12},
and the second in which gravity and the background
magnetic field are parallel to each other \citep{bra+gre+jab+kle+rog14,los+bra+kel+rog14},
the analysis of the instability simplifies.
Unfortunately, the problem is more complicated in the present case where all the
three components of magnetic field are present.
In that case, a systematic determination of the new transport coefficients in the
effective magnetic pressure, using direct numerical simulations (DNS),
has not yet been performed.
Nevertheless there are two signatures of NEMPI that we look for.
Firstly, we know the effective magnetic pressure is {\it negative}
only when the background magnetic field is neither too large or too small,
within $0.1$ to $1$ when normalized by the equipartition magnetic field
\citep{bra+kem+kle+rog12}.
We find that this condition is satisfied near the top surface when the first flux
emergence occurs, as shown in \subfig{fig:beq_diff_times}{a}, but not at later stages
as shown in \subfig{fig:beq_diff_times}{b}.
What is then the mechanism behind the disappearance and reappearance of the magnetic flux
at the top surface?
A clue to this puzzle is the fact that within mean-field theory the dynamo operating
in the lower layers of the computational domain can be interpreted as an $\alpha^2$
dynamo, where
$\alpha \propto -\Tto\brah{\ww\cdot\uu}$,
where $\ww=\WW-\Wbar$ and $\uu=\UU-\Ubar$ are fluctuations.
An $\alpha^2$ dynamo for which $\alpha$ varies within the domain can give rise
to dynamo waves~\citep{bar+shu87,ste+ger04,mit+tav+kap+bra09}, and indeed such dynamo
waves are seen in our simulations as shown in the space-time diagram in \fig{fig:bxby_bfly}.

\begin{figure}
\iftoggle{psfig}{
  \includegraphics[width=0.85\columnwidth]{fig/runA_pBtop.ps}
  \includegraphics[width=0.85\columnwidth]{fig/runA_pUtop.ps} \\
  \includegraphics[width=0.85\columnwidth]{fig/runA_pUtop_Ffilter.ps}
  \includegraphics[width=0.9\columnwidth]{fig/runA_Ufilter_diag2d.ps}
}{
  \includegraphics[width=0.85\columnwidth]{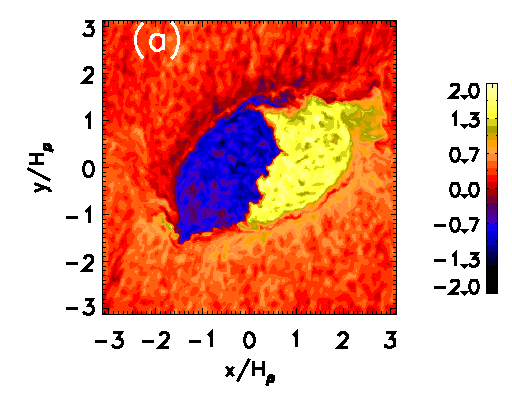}
  \includegraphics[width=0.85\columnwidth]{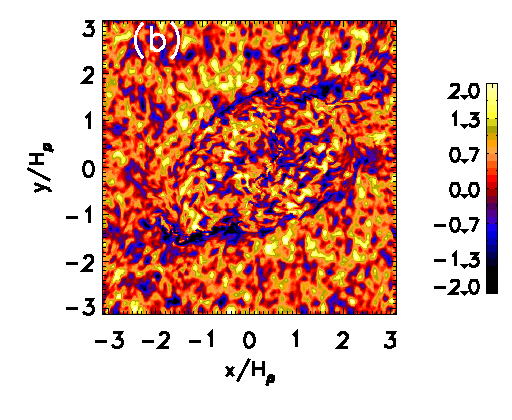} \\
  \includegraphics[width=0.85\columnwidth]{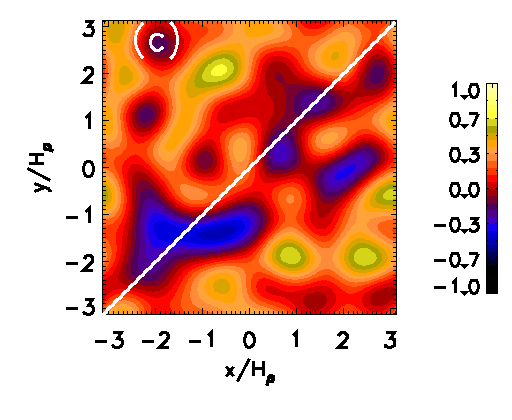}
  \includegraphics[width=0.9\columnwidth]{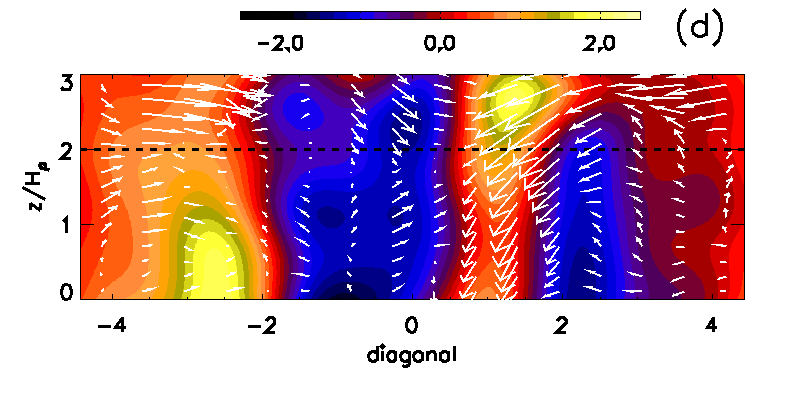}
}
\caption{(a) Contour plot of $B_z/\Beqz$ at $z=3$ from {\tt Run~A}.
(b) $U_z/\urms$ at $z=3$ from the same snapshot.
(c) The $U_z/\urms$ Fourier filtered by a low-pass filter with Fourier mode $\kappa=5$, $\braf{U_z}$,
from the same snapshot. The white line shows the diagonal.
(d) The flow velocities  (Fourier filtered) as arrows in the vertical plane along the
diagonal plotted in (c).  The pseudo-colors represent  $B_z/\Beqz$.
The in-plane components are plotted as arrows.
Note the downward flow near the top surface.
}
\label{fig:fourier_flow}
\end{figure}

The second signature of NEMPI is its ability to generate large-scale flows;
since NEMPI creates regions of negative effective magnetic pressure, it is often
accompanied by a converging flow at the surface and a downward flow on and immediately below
the location of flux concentration\footnote{In general converging flows are typically observed in simulations of
stratified convection.
Such flows can be  quite effective in concentrating vertical magnetic
flux.
The crucial input coming from the concept of  NEMPI is that the converging
flows themselves are generated by NEMPI due to the presence of weak background
magnetic field.}.
In our simulations, due to the presence of strong turbulent
fluctuations, we have not been able to detect any such coherent flow, although
some evidence in support of such a flow has been found in the Fourier filtered velocity
field as shown in \fig{fig:fourier_flow}.
Interestingly, similar downflows are also seen in recent simulations
by \cite{rem+che14}, who inject a $10 \rm{kG}$ flux tube at the bottom of a
solar convection simulation and let it rise to the surface.
Although the emergence process itself is associated with upflows, their results show downflows
at the late stages of the flux concentration process.
In such simulations that attempt to be realistic, it is
not possible to attribute the observed downflows to one single mechanism.
By contrast, in our simple setup it is likely that NEMPI is
indeed the mechanism responsible for generating the downward flow.

\section{Conclusion}

To conclude, in this paper, we have shown that it is possible
to generate intense structures of vertical magnetic field
at the top surface of DNS of a density-stratified turbulent dynamo.
Furthermore, a rich dynamic behaviour of the magnetic field
is observed: bipolar spot-like structures appear, then morph into
bipolar band-like structures which
disappear and reappear at a different place and at a later stage
evolve into spot-like structures.
Such structures are similar to $\delta$ spots \citep[see, e.g.,][]{fisher}
and tend to show anticlockwise rotation, which is consistent with the
fact that the kinetic helicity in our simulations is positive.

The characteristic length and time scales of the magnetic field
formed at the top surface are much smaller than the characteristic length scale (and time scale) of
the dynamo-generated magnetic field.
The necessary conditions are strong stratification, presence of turbulence,
and large scale separation, which is at least $30$ in the DNS we present here.
Clearly, there is a mechanism at work here that can concentrate a weak
large-scale magnetic field to strong magnetic field of smaller scale.
Could this mechanism be NEMPI?
At present, we cannot provide a definitive answer to this question,
although we do show that the necessary conditions for NEMPI to operate
are satisfied during the first emergence of flux at the top surface.

How relevant are our result in understanding the formation of active regions and
sunspots?
Unlike the works by e.g., \cite{ste+nor12} or \cite{rem+che14}, our simulations
do not include radiative hydrodynamic convection; turbulence is generated
by external forcing.
This should not necessarily be considered a shortcoming of our simulations
as the aim of our work has been to present the simplest model that can show
formation of bipolar structures from a large-scale dynamo.
This is the first time bipolar structures are found to appear
in simulations where the magnetic field is not imposed -- as is the
case in \cite{ste+nor12}, \cite{war+los+bra+kle+rog13}, or \cite{rem+che14}
-- but it is self-consistently generated from
a dynamo in strongly stratified forced turbulence.

The most remarkable feature of these simulations is that a minimalistic
setup consisting solely of stratification and helically forced turbulence
can generate such diverse spatio-temporal behaviour.
Could a mean-field model consisting of both dynamo equations
and equations describing NEMPI capture such behaviour?
This question will be the subject of future investigations.

\section*{Acknowledgements}

Financial support from the European Research Council under the AstroDyn
Research Project 227952, the Swedish Research Council under the grants
621-2011-5076 and 2012-5797, the Research Council of Norway
under the FRINATEK grant 231444,
as well as from the Government of the Russian Federation under
a grant 11.G34.31.0048
are gratefully acknowledged.
The computations have been carried out at the National Supercomputer
Centres in Link\"oping and Ume{\aa} as well as the Center for Parallel
Computers at the Royal Institute of Technology in Sweden and the Nordic
High Performance Computing Center in Iceland.

\end{document}